# The number system hidden inside the Boolean satisfiability problem

**Author:** Keum-Bae Cho

**Affiliations**

Department of Electrical and Computer Engineering, Seoul National University, 1 Gwanak-ro, Gwanak-gu, Seoul, 151-742, South Korea

*Author to whom correspondence should be addressed: E-mail (kbcho98@snu.ac.kr)

**Abstract**

This paper gives a novel approach to analyze SAT problem more deeply. First, I define new elements of Boolean formula such as dominant variable, decision chain, and chain coupler. Through the analysis of the SAT problem using the elements, I prove that we can construct a $k$-SAT ($k>2$) instance where the coefficients of cutting planes take exponentially large values in the input size. This exponential property is caused by the number system formed from the calculation of coefficients. In addition, I show that 2-SAT does not form the number system and Horn-SAT partially forms the number system according to the feasible value of the dominant variable. Whether or not the coefficients of cutting planes in cutting plane proof are polynomially bounded was open problem. Many researchers believed that cutting plane proofs with large coefficients are highly non-intuitive[20]. However, we can construct a $k$-SAT ($k>2$) instance in which cutting planes take exponentially large coefficients by the number system. In addition, this exponential property is so strong that it gives definite answers for several questions: why Horn-SAT has the intermediate property between 2-SAT and 3-SAT; why random-SAT is so easy; and why $k$-SAT ($k>2$) cannot be solved with the linear programming technique. As we know, 2-SAT is NL-complete, Horn-SAT is P-complete, and $k$-SAT ($k>2$) is NP-complete. In terms of computational complexity, this paper gives a clear mathematical property by which SAT problems in three different classes are distinguished. Two questions, NL =? P and P =? NP, have been open problems for several decades. This study presents a definite supporting evidence for the conjecture that NL $\subsetneq$ P $\subsetneq$ NP and a new solving direction for the P versus NP problem.

# 1. Introduction

The class P (polynomial) is the set of decision problems that are solvable in polynomial time and the class NP (nondeterministic polynomial) is the set of decision problems for which a solution can be verified in polynomial time. More precisely, a polynomial time algorithm for a problem is a method that solves the problem correctly on every input and takes no more than $c \cdot n^k$ time on an input of size $n$ (i.e., $O(n^k)$) for some constants $c>0$ and $k>0$. NP is the class of languages that have a polynomial-time verification algorithm. Any deterministic Turing machine can be simulated by a non-deterministic Turing machine with no overhead. Thus, P is included in NP. Then, if a decision problem can be verified in polynomial time, is it possible to solve the problem in polynomial time? This question is the well-known P versus NP problem[1], which is to clarify the relationship for the inclusion of the classes P and NP. Cook and Levin proposed the question about forty years ago as a problem concerned with the fundamental limits of feasible computation[2,3]. The NP-complete problem is a problem that belongs to NP and every problem in NP is reducible to the problem in polynomial time. The obvious way to prove P = NP is to show that some NP-complete problem has a polynomial time algorithm. Researchers have found thousands of NP-complete problems since Karp's research[4,5]. However, although there are so many NP-complete problems, researchers have failed to find a polynomial time algorithm for any one of the problems. Hence, various proof techniques have been studied to distinguish between P and NP with the belief that $P \neq NP$. However, all known proof techniques such as relativizing, natural, and algebrizing proofs are insufficient to prove that $P \neq NP$. Baker, Gill, and Solovay showed that P = NP with respect to some oracles, while $P \neq NP$ for other oracles. Hence, the P versus NP question cannot be solved by any of the proof techniques that separate complexity classes relative to an oracle[6]. Every polynomial-time computable function can be expressed by a circuit with a polynomial number of gates[7]. From this relationship, exponential lower bounds have been proved for restricted circuit models such as monotone circuits[8,9] and bounded depth circuits with unbounded fan-in gates[10,11]. However, Razborov and Rudich proved that if one-way functions exist, no natural proof method could distinguish between P and NP[12]. Although one-way functions have never been formally proven to exist, most researchers believe that a proof or disproof of the existence of a one-way function would be much harder than clarifying the relationship of the classes P and NP. An algebrizing proof was successfully used to prove several complexity theories such as IP = PSPACE[13,14] and PCP theorem[15]. However, Aaronson and Wigderson showed that algebrizing technique is fundamentally unable to resolve the barrier problem of P versus NP. They pointed out that the reason for the incapability to solve this problem is the failure of opening the Boolean formula wide enough. Thus, it needs to probe the Boolean formula in some deeper way for further progress[16].

In this paper, we introduce a novel idea to analyze the Boolean formula more deeply. The Boolean or propositional satisfiability (SAT) problem is to determine whether there exists a feasible set to satisfy a given Boolean formula. SAT is the first known example of a NP-complete problem and thousands of NP-compete problems have been identified by reducing the SAT to the NP-complete problems. There are several special cases of satisfiability problem. The *k*-SAT determines the satisfiability of a Boolean formula in conjunctive normal form (CNF) where each clause is limited to at most *k* literals. Especially, 3-SAT is contained in the 21 NP-complete problems researched by Karp. Class NL (nondeterministic logarithm) consists of the decision problems that can be solved by a nondeterministic Turing machine with a read-only input tape and a separate read-write tape whose size is limited to be proportional to the logarithm of the input length. The NL-complete problem is defined as a decision problem where the problem belongs to NL and has the additional property that every other decision problem in NL can be reduced to the NL-complete problem. 2-SAT belongs to the NL-complete problem[17]. Similarly, the P-complete problem is defined as a decision problem included in P and every problem in P can be reduced to the P-complete problem by using an appropriate reduction. Horn-SAT belongs to the P-complete problem[18], which consists of Horn clauses that contains at most one positive literal. XOR-SAT is another special case of the SAT where each clause contains exclusive OR operators rather than the OR operators. XOR-SAT belongs to P since an XOR-SAT formula can be solved in cubic time by Gaussian elimination. There are only six tractable (polynomial time decidable) cases in SAT: 2-SAT, Horn-SAT, dual-Horn-SAT, XOR-SAT, instances satisfied by the all '0' assignment and instances satisfied by the all '1' assignment. Otherwise, the SAT problem can be reduced by 3-SAT. Therefore, SAT is polynomial time decidable or NP-complete because 3-SAT is NP-complete. This relationship was termed the Schaefer Dichotomy Theorem[19]. However, we do not know whether 3-SAT can be polynomial-time decidable up to now. As shown above, SAT contains NL-complete, P-complete, and NP-complete problems. Thus, SAT is a good research subject to search for some intrinsic property that appears only in NP-complete problems. There have been no ideas for a deep analysis of the SAT structure. We newly define a dominant variable, decision chain, and chain coupler based on the characteristics of the SAT. Through the analysis of SAT structure using the dominant variable, decision chain, and chain coupler, we derive the natural number system hidden inside $k$-SAT ($k>2$). In addition, we show that the number system is not formed in 2-SAT, but partially formed in Horn-SAT according to the feasible value of a dominant variable, and always formed in $k$-SAT ($k>2$) regardless of the feasible value of a dominant variable. Thus, this study gives us clear answer for the following research question: Is there any definite mathematical expression to explain why Horn-SAT has an intermediate property between 2-SAT and 3-SAT?

## 2. Definitions

Suppose that a Boolean formula in a conjunctive normal form (CNF) is given:

$$\Phi(X) = (X_1 \vee \neg X_2 \vee \neg X_4) \wedge (X_1 \vee \neg X_3) \wedge (X_2 \vee X_3) \wedge (\neg X_2 \vee X_4) \cdots (1)$$

We need some sort of measure to say that a problem is easy or hard. It is natural to expect this measure to be represented with a number indicating a definite physical meaning. In order to search for the measure in geometric relationships, we reduce a given SAT instance to a 0-1 integer-programming instance. The procedure to systematically reduce a SAT instance to an equivalent 0-1 integer programming instance is as follows:

First, we change the literals such as 'X' and '¬X' to variables such as '$x$' and '$1-x$', and a logical operator '∨' to an arithmetic operator '+'. Second, we assign a lower bound and an upper bound to represent the constraints of a clause with inequities. The lower bound is '1' due to the satisfiability constraint of the clause. The upper bound is the number of literals in the clause due to the 0-1 integral constraint of input variables. Third, we merge all generated inequalities using an integral matrix and integral vectors. For instance, from eq. (1),

$1 \leq x_1 + (1-x_2) + (1-x_4) \leq 3, \ 1 \leq x_1 + (1-x_3) \leq 2, \ 1 \leq x_2 + x_3 \leq 2, \ 1 \leq (1-x_2) + x_4 \leq 2$

$-1 \leq x_1 - x_2 - x_4 \leq 1, \ 0 \leq x_1 - x_3 \leq 1, \ 1 \leq x_2 + x_3 \leq 2, \ 0 \leq -x_2 + x_4 \leq 1$

$$\begin{bmatrix} -1 \\ 0 \\ 1 \\ 0 \end{bmatrix} \leq \begin{bmatrix} 1 & -1 & 0 & -1 \\ 1 & 0 & -1 & 0 \\ 0 & 1 & 1 & 0 \\ 0 & -1 & 0 & 1 \end{bmatrix} \begin{bmatrix} x_1 \\ x_2 \\ x_3 \\ x_4 \end{bmatrix} \leq \begin{bmatrix} 1 \\ 1 \\ 2 \\ 1 \end{bmatrix}, \quad x_i \in [0,1] \quad \cdots(2)$$

Note that, we relaxed the integral constraint of variables to use linear programming techniques in eq. (2) (Linear programming relaxation). If the integral matrix is totally unimodular (the determinant of every square sub-matrix is 0, +1, -1), the polytope is integral (each of its nonempty faces contains an integral point). Then, we can acquire the solution set using linear programming techniques. However, if not, we have no polynomial time algorithm to solve the problem yet. In addition, if the polytope is not integral, it means that some of the infeasible integral points are located near the polytope. As the distance is smaller, we must search for a separation hyperplane passing through a smaller area to separate the infeasible integral point from the polytope. Hence, it is reasonable that the measure for the hardness of the SAT instance is the size of the distance from infeasible integral points to the polytope. Based on this concept, our interest is to verify how small the distance is and to search for how to make the distance smaller.

In this paper we mainly use two techniques: resolution and cutting planes generation. If two clauses have exactly one conflicting literal, they produce a new clause implied by two clauses. The resolution rule is a single valid inference rule, which is written by:

$$\frac{\left(\vee_{k=1}^{i} a_k\right) \vee c \quad \left(\vee_{k=1}^{j} b_k\right) \vee \neg c}{\left(\vee_{k=1}^{i} a_k\right) \vee \left(\vee_{k=1}^{j} b_k\right)} \quad \cdots(3)$$

Cutting planes are inequalities of the form where the coefficients and constant terms are integers, and the variables are Boolean variables, which generates from earlier linear inequalities by the linear combination rule or the cut rule. These are respectively,

$$\frac{\sum \lambda_i^1 x_i \geq t_1 \quad \cdots \quad \sum \lambda_i^k x_i \geq t_k}{\sum \left(\sum_j s_j \lambda_i^j\right) x_i \geq \sum_j s_j t_j} \quad \text{and} \quad \frac{\sum s \lambda_i x_i \geq t}{\sum \lambda_i x_i \geq \lceil t/s \rceil} \quad \cdots(4)$$

In eq. (4), $s_1$, ..., $s_k$ and $s$ must be strictly positive integers. All cutting planes represent necessary and unnecessary constraints. If the inequality of a cutting plane represents a necessary constraint to generate the polytope, it is called as a facet-defining inequality and mapped to a facet. In this paper, we focus on the normal vector of the cutting planes, and thus, we do not care about round process in the cut rule. Our objective is to make a cutting plane extremely near the integral point using the characteristics of the SAT problem. Hereafter, we use CNF mixed with a matrix in eq. (2) and we use positive variable, negative variable mixed with literal, and negation of the literal.

2.1 Dominant variable, decision margin

A *dominant variable* is defined as a variable that takes only one value between '0' and '1' in the solution set. If a variable is the dominant variable, we say that the variable has a *dominance* property. An *infeasible point of a dominant variable* is defined as an integral point where the value of the dominant variable is not feasible.

The number of infeasible points is $2^{n-1}$ in *n*-dimensional space. We can draw $2^{n-1}$ lines that pass through one of the infeasible points and parallel to the dominant variable axis. These lines are termed as ***decision lines***. The intersection point between a decision line and a facet of the polytope is termed a ***decision point***. If we want to know whether a variable is a dominant variable or not, we need to verify that all $2^{n-1}$ infeasible points are located outside the polytope in *n*-dimensional space. Therefore, if we want to verify in polynomial time, we must reduce the dimension by eliminating the related variables in terms of a linear system or projecting the polytope to lower dimensional space in terms of geometry.

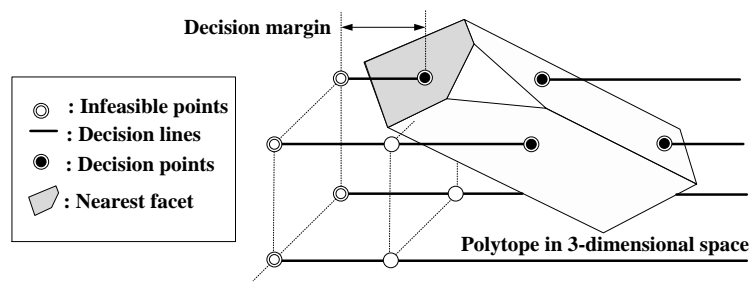

**Fig. 1.** The concept of the decision margin

The smallest distance from the infeasible point to the decision point in *r*-dimensional space is termed as a ***decision margin in r-dimensional space***. Figure 1 describes the concept of the decision margin. The decision margin becomes the measure for the hardness of a SAT instance. If a SAT instance is unsatisfiable, then at least one variable is a dominant variable of which the feasible value is '0' in some set of clauses and of which the feasible value is '1' in another set of clauses. Thus, we must verify that the total $2^n$ numbers of infeasible points are outside the polytope.

2.2 Decision chain, chain coupler

Suppose that we synthesize a SAT instance containing a dominant variable. If we remove the dominant variable from all clauses including the variable, the new generated CNF should be unsatisfiable. Otherwise, the removed dominant variable can be assigned with any value of both '0' and '1' as a feasible value. This contradicts the definition of the dominant variable. Hence, we first synthesize an unsatisfiable CNF and then insert a dominant variable to make a SAT instance containing a dominant variable. One of the simplest ways to make an unsatisfiable CNF is to use the resolution technique. The conjunctions of implication chains such as $(X_1 \rightarrow X_2) \cap (X_2 \rightarrow X_3) \cap \ldots \cap (X_{k-1} \rightarrow X_k)$ is reduced $(\neg X_1 \vee X_k)$ by the resolution steps. Thus, we can easily verify that $X_1 \cap (X_1 \rightarrow X_2) \cap (X_2 \rightarrow X_3) \cap \ldots \cap (X_{k-1} \rightarrow X_k) \cap \neg X_k$ is unsatisfiable, which is represented as $X_1 \wedge (\neg X_1 \vee X_2) \wedge (\neg X_2 \vee X_3) \wedge \ldots \wedge (\neg X_{k-1} \vee X_k) \wedge \neg X_k$ in a CNF. Then, we modify the clauses in remaining the unsatisfiablilty.

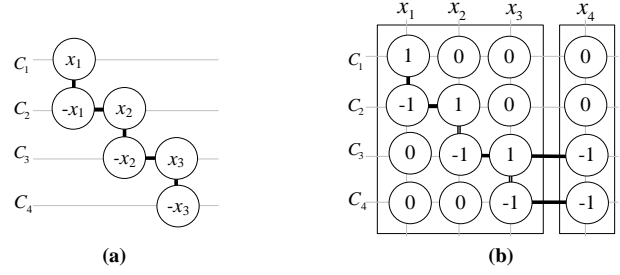

**(a)**            **(b)**

**Fig. 2.** An instance of a decision chain and dominant variable

Figure 2(a) shows a chain representation of an unsatisfiable CNF corresponding to $X_1 \wedge (\neg X_1 \vee X_2) \wedge (\neg X_2 \vee X_3) \wedge \neg X_3$. Figure 2(b) shows the matrix and vector representation of the unsatisfiable CNF and a dominant variable connected to the CNF. Let us think about the method to determine feasible values when a dominant variable is connected to an unsatisfiable CNF. We assign a TRUE value ('1' when the coefficient is '1', '0' when the coefficient is '-1') to the dominant variable. Then the clauses containing the dominant variable are satisfiable regardless of the values of the other variables, which makes it possible to assign a feasible value to a variable in a neighboring clause to make the clause satisfiable. Then, we can assign a feasible value to a variable in another neighboring clause to make the clause satisfiable. If we repeat the above process, all variables are assigned with feasible values. This decision process is compared to the ignition of fire. If the number of dominant variables connected to the unsatisfiable CNF is two or more, we can take another dominant variable as an igniter. Then we can obtain another solution set. We term this unsatisfiable clause set as a ***decision chain*** because the decision is executed through the connected lines like a chain. As mentioned above, we need to eliminate variables to reduce the number of infeasible points for the polynomial time algorithm. For easy elimination, we add a constraint to the decision chain that ***all variables should be eliminated with addition of all clauses in the decision chain.*** The above instance satisfies this condition.

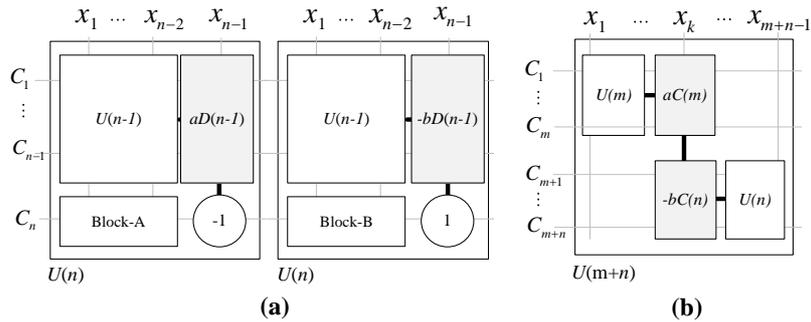

**(a)**            **(b)**

**Fig. 3**. New decision chain generation with a dominant variable and chain coupler

A dominant variable can be connected to two decision chains with a different sign. We term the dominant variable connecting two decision chains as a ***chain coupler***. Every dominant variable and chain coupler can be connected in multiple times to a decision chain. However, they must have the same sign in all clauses contained in a decision chain. Otherwise, the variable always satisfies at least one clause regardless of the value of the variable and then, the decision chain is satisfiable. This contradicts the definition of a dominant variable. Figure 3(a) shows a newly generated decision chain by a dominant variable. Block-A and Block-B can be assigned by arbitrary coefficients in remaining unsatisfiablilty. Figure 3(b) shows a newly generated decision chain by a chain coupler $x_k$. The decision chain, which is an **U**nsatisfiable clause set with $m$ columns (the number of clauses) and $n$ rows (the number of variables), is denoted by matrix notation U($m\times n$). Every decision chain used in this paper satisfies the relation $m=n+1$. Hence, we simply express with **U(m)** instead of U($m\times n$). The chain **C**oupler and **D**ominant variables are denoted by column vector notation $\pm k\mathbf{C}(m)$ and $\pm k\mathbf{D}(m)$, respectively, where $m$ is the number of columns and $k$ is the magnitude of the sum of all coefficients connected to a decision chain. Column and row exchanging does not affect the satisfiability. Thus, we can exchange columns and rows to make a decision chain in a CNF matrix.

## 3. SAT structure analysis

Our interest is to acquire exact inequalities of the nearest facet in $r$-dimensional space to calculate the decision margin. Acquiring exact inequalities is attended by the variable reduction process. Hence, we will survey how to construct a CNF in order to easily reduce the variables. Consider when two or more variables are connected to a decision chain. Figure 4(a) describes a decision chain connected by $m+2$ variables in 3-SAT.

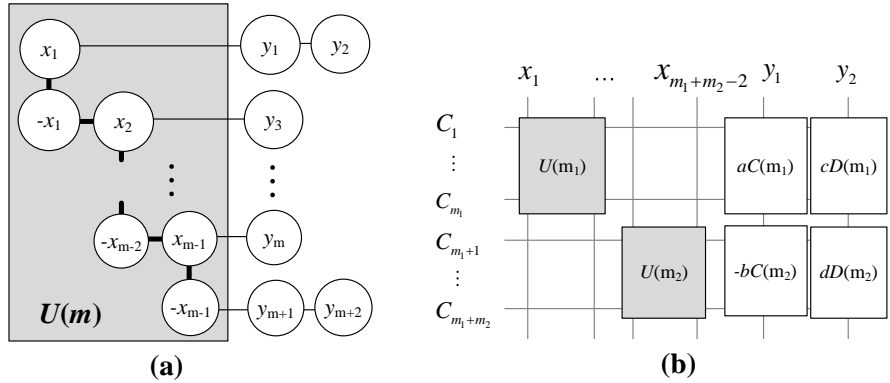

**Fig. 4**. The structures of SAT composed of a decision chain and dominant variable candidates: (a) dominant variable candidates connected to a decision chain in 3-SAT; and (b) two decision chains connected by variables $y_1$ and $y_2$

We can easily verify that decision chain $U(m)$ can be connected by at most $m(k$-$2)+2$ number of variables in $k$-SAT ($k \geq 2$). Note that $y_1$ has a different sign in two decision chains and $y_2$ has the same sign in Fig. 4(b). The variable $y_2$ must take only '1' to satisfy all clauses. Thus, $y_2$ becomes a dominant variable and $y_1$ becomes a chain coupler. If $a$ and $b$ are all '1', two decision chains are combined to a new larger decision chain.

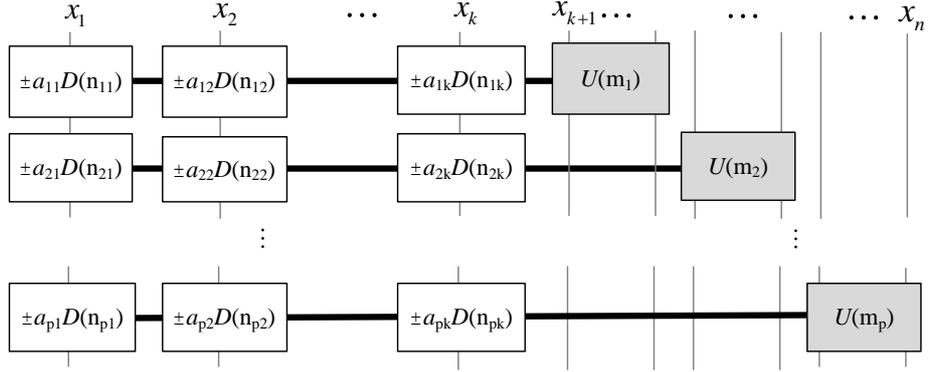

**Fig. 5**. Parallel connected dominant variable candidates in multiple decision chains

Figure 5 shows *p* number of decision chains and *k* number of dominant variable candidates. If a variable in a decision chain takes the TRUE value then the other variables can take any value of '0' or '1'. For this reason, we added the term 'candidates'. Figure 5 shows another SAT instance existing inside a given SAT instance. At least one variable must be assigned with the TRUE value to satisfy all clauses constructing a decision chain. This condition should be satisfied in all decision chains. Thus, every decision chain generates a new clause consisting of variables $x_i$ ($1 \leq i \leq k$), and the new generated clauses are combined with the AND constraints. This process makes a new CNF. The SAT of $\{x_1, \ldots, x_n\}$ is satisfiable if and only if the SAT of $\{x_1, \ldots, x_k\}$ is satisfiable. We can enclose a small CNF with a larger CNF. This characteristic will be used for the hard SAT generation algorithm. Now, let us investigate the structure of a CNF in terms of the dominant variable candidate, decision chain, and chain coupler. Dominant variable candidates have the same coefficient in all decision chains, and chain couplers have different coefficients in two decision chains in remaining the same sign inside a decision chain. A dominant variable can be connected one or more times to a decision chain. In addition, we can connect a string of dominant variable candidates to a decision chain. Figure 6 shows a sub-matrix of a SAT instance constructed with *k* number of dominant variable candidates, *e* number of decision chains using *n-e-k+1* number of variables, and *e*-1 number of chain couplers.

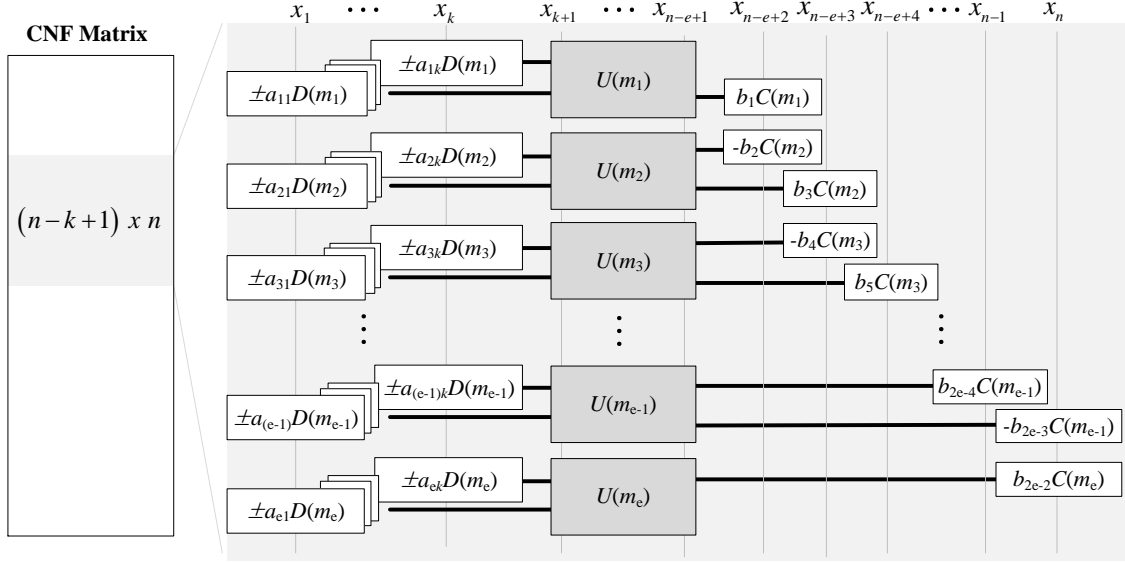

**Fig. 6.** The structure of a sub-matrix constructed with dominant variable candidates, decision chains, and chain couplers

We will show that the coefficients of the dominant variable candidates in inequalities of cutting planes can be expressed by the natural number system of which the exponent is exponential in the input size. We bind decision chains and chain couplers in Figure 6 with one block. This block is termed as a **decision block.** A decision block is another decision chain larger than a decision chain inside the decision block. We added a constraint to the decision chain that all variables should be eliminated with the addition of all clauses. We can make new inequalities that remain only dominant variable candidates $x_i$ ($1 \leq i \leq k$) using the addition and multiplication process in Figure 6. First, we want to remove all variables in $U(m_1)$, $U(m_2)$, and $x_{n-e+2}$ at the same time. We multiply all clauses contained in $U(m_1)$ by $b_2$ and all clauses contained in $U(m_2)$ by $b_1$. We then add all clauses contained in the decision chains $U(m_1)$ and $U(m_2)$. By the added constraint of the decision chain, if we add all clauses constructing the decision chain at the same time, all variables constructing the decision chain are erased. In this stage, variable $x_{n-e+3}$ is only remained. Second, we multiply all clauses that remain in $U(m_2)$ by $b_4$ and all clauses that are contained in $U(m_3)$ by $b_1 b_3$ to remove the variable $x_{n-e+3}$. The value $b_1 b_3$ indicates the multiplier of the chain coupler that is generated during the previous elimination step. We then add all clauses that are contained in the decision chains $U(m_2)$ and $U(m_3)$. Then, $U(m_3)$ and $x_{n-e+3}$ are removed and only $x_{n-e+4}$ remains. If we repeat this process, all variables constructing the decision chains and chain couplers are erased, and newly generated inequalities are expressed as:

$$c_{\min} \leq \sum_{j=1}^{k} a_j x_j \leq c_{\max} \; (c_{\min}, c_{\max} \in \mathbb{Z}) \Big| a_j = \Big( \cdots \big((a_{1j}b_2 + a_{2j}b_1)b_4 + a_{3j}b_1 b_3\big)b_6 + a_{4j}b_1 b_3 b_5 \cdots \Big) b_{2e-2} + a_{ej} b_1 b_3 b_5 \cdots b_{2e-3} \quad \cdots (5)$$

# 4. Number system in coefficients

**Lemma 1. The coefficient of the facet can be expressed by the number system of which the exponent is exponential in the input size in *k*-SAT (*k* >2).**

*Proof.* In eq. (5), if we assign '1' to $b_i$ where $i$ is an odd number, and we assign $b$ to $b_i$ where $i$ is an even number, then the coefficients of $x_j$ ($1 \leq j \leq k$) is expressed by the number system with basis $b$ as:

$$a_j = \sum_{i=1}^{e} a_{ij} b^{i-1} \quad \cdots (6)$$

The maximum number of variables that can be connected to a decision chain containing $c$ clauses is $c+2$ in 3-SAT. The relation of $b$, $c$, and multipliers of dominant variable candidates in 3-SAT is:

$$A_j + b \leq c + 2 \bigg| A_j = \sum_{i=1}^{k} a_{ji}, j \in \{1, e\}, \quad A_j + b \leq c + 1 \bigg| A_j = \sum_{i=1}^{k} a_{ji}, 2 \leq j \leq e-1 \quad \cdots (7)$$

The chain coupler can be connected to a clause only one time, and thus, $b \leq c$.

In Figure 6, the relation of the number of variables is calculated as:

$$n = ce + (e-1) + d \quad \cdots (8)$$

In eq. (8), $n$ is the input size, $c$ is the number of variables forming the decision chains, $e$ is the number of decision chains, and $d$ is the number of dominant variable candidates.

From eq. (6) and eq. (8),

$$a_j = \sum_{i=1}^{\left\lfloor \frac{n-d+1}{c+1} \right\rfloor} a_{ij} b^{i-1} \quad \cdots (9)$$

In eq. (9), symbol $\lfloor x \rfloor$ means the largest integer that does not exceed $x$. This value is exponential on $n$ because $c$ and $d$ are constants, and $b$ can be assigned with any value satisfying eq. (7). ∎

For example, if we assign the number of dominant variables with '1', the maximum value of $b$ becomes $c+1$ in 3-SAT. The number $c+1$ means the number of clauses in a decision chain. If we assign '1' to all $a_{ij}$ in eq. (6), the maximum value of the coefficient is represented as:

$$\frac{c^{\left\lfloor \frac{n}{c} \right\rfloor} - 1}{c - 1} \quad (c \geq 2, c \in \mathbb{N}) \quad \cdots (10)$$

The number $c$ means the number of clauses in a decision chain in eq. (10).

# 5. Exponential property of decision margin

**Lemma 2. If the coefficient of the facet takes an exponentially large value, the decision margin decreases to an exponentially small value.**

*Proof.* We prove in 2-dimensional space without loss of generality. Then, all variables are '0' except for $x_1$ and $x_2$ in eq. (5). We can change eq. (5):

$$c_{\min} \leq a_1 x_1 + a_2 x_2 \leq c_{\max} \,\Big|\, a_j = \sum_{i=1}^{\left\lfloor \frac{n-d+1}{c+1} \right\rfloor} a_{ij} b^{i-1}, j \in \{1,2\} \quad \cdots(11)$$

We do not know the value of $c_{\min}$ and $c_{\max}$ in eq. (11). Thus, we cannot easily confirm an exponentially small decision margin with only the range of the coefficients, $a_1$ and $a_2$, because $c_{\min}$ and $c_{\max}$ can be an exponentially large value. However, we can solve this problem by comparing the feasible range of a variable in different decision lines. Figure 7 shows an instance that added clauses make two infeasible points of $x_1$ of which the value is '0'.

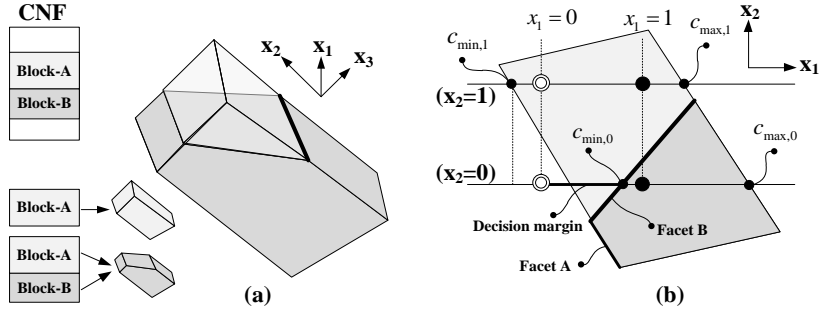

**Fig. 7.** The feasible region of $x_1$: (a) projected polytope in 3-dimensional space; and (b) feasible region of $x_1$ in the $x_1$-$x_2$ plane

Suppose that there is a CNF instance (Block A) where $x_1$ takes both '0' and '1' as a feasible value. In this stage, we add a sub-set of clauses (Block B) to the CNF. To add clauses in a CNF means to add the constraints in terms of a linear system and to generate new cutting planes in terms of geometry. Suppose that these constraints make $x_1$ take only one value of '1'. If we assign the FALSE value ('0' when the coefficient is '1', '1' when the coefficient is '-1') to $x_2$, then $x_1$ should be a dominant variable of which the feasible value is '1'. However, if we assign the TRUE value to $x_2$, then $x_1$ cannot be a dominant variable anymore and $x_1$ can take both '0' and '1' as a feasible value. Suppose $a_2$ is a positive integer. Then, we can calculate $c_{\min,0}$ from the inequality (11) by adding the condition that $x_2=0$ and $c_{\min,1}$ by adding the condition that $x_2=1$

$$c_{\min,0} = \frac{c_{\min}}{a_1}, \quad c_{\min,1} = \frac{c_{\min} - a_2}{a_1} \quad \cdots(12)$$

The cutting line inside a parallelogram is generated from the inequalities of the added clauses. We can use the characteristic of a parallelogram because the projection is an affine transformation. The feasible region of the decision line where $x_2=0$ must not include the integral point where $x_1=0$. In addition, the feasible region of the decision line where $x_2=1$ must include the integral point where $x_1=0$. Therefore, the line $x_1=0$ must locate between $c_{min,0}$ and $c_{min,1}$. The difference of $c_{min,0}$ and $c_{min,1}$ is calculated to $a_2/a_1$ from eq. (12). Therefore, the decision margin must be less than or equal to $a_2/a_1$. We can make $a_1$ any number of exponential size and $a_2$ any number of polynomial size by Lemma 1. As a result, if the coefficient of the facet takes an exponentially large value, the decision margin decreases to an exponentially small value. ∎

## 6. Number system in tractable SAT

Horn-SAT is the one of the hardest problems among the tractable SATs in the sense that it is a P-complete problem[18]. We know that 2-SAT is NL-complete, Horn-SAT is P-complete, and 3-SAT is NP-complete. In addition, we predict the relations of NL, P, and NP as NL⊊ P ⊊ NP. Hence, we expect that Horn-SAT has an intermediate characteristic between 2-SAT and 3-SAT. Then, is there any definite mathematical expression to explain why Horn-SAT has an intermediate property between 2-SAT and 3-SAT? The following lemma gives us the answer for this question.

**Lemma 3. 2-SAT does not form the number system and Horn-SAT forms the number system only when the feasible value of the dominant variable is '0'.**

*Proof.* The number system is formed by the role of the chain coupler. Thus, if we cannot make multiple-connected chain couplers, the number system cannot be formed. A decision chain in 2-SAT is connected by at most two variables as shown in Figure 8(a). Thus, all chain couplers can be connected only once to a decision chain as shown in Figure 8(b) since at least one variable should be used for a dominant variable or another chain coupler. As a result, the number system is not formed because exponent $b$ becomes '1' in eq. (9).

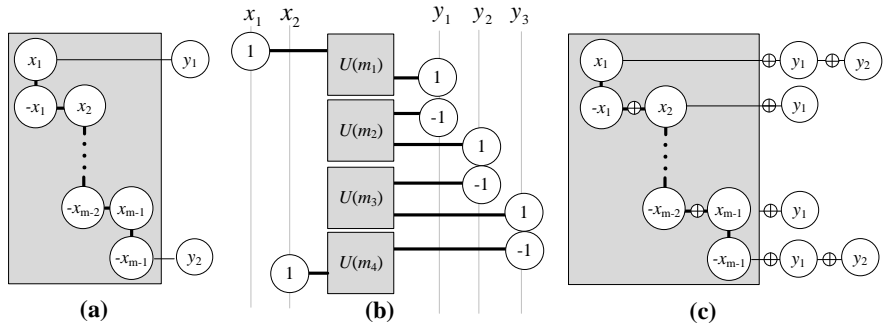

**Fig. 8.** 2-SAT and XOR-SAT structure

If we assign the TRUE value to a variable in a clause, the clause is always satisfiable regardless of the other variable's values in normal SAT. Therefore, two or more connections of a dominant variable in a decision chain do not affect the satisfiability of the clause. However, the concept of the multiple-connection is not valid in XOR-SAT. We can make a decision chain with XOR-SAT like a normal SAT as shown in Figure 8(c). However, if we simultaneously add the same variables in two or more clauses contained in a decision chain, the number of multiple connections affects the satisfiability of the decision chain. As a result, the number system cannot be formed in XOR-SAT because we cannot make a multiple-connected chain coupler and multiple-connected dominant variable candidates.

Horn-SAT has at most one variable with a positive coefficient in all clauses. Due to this constraint, only one positive variable can be connected to a decision chain. This positive variable can be used as a dominant variable candidate or a chain coupler.

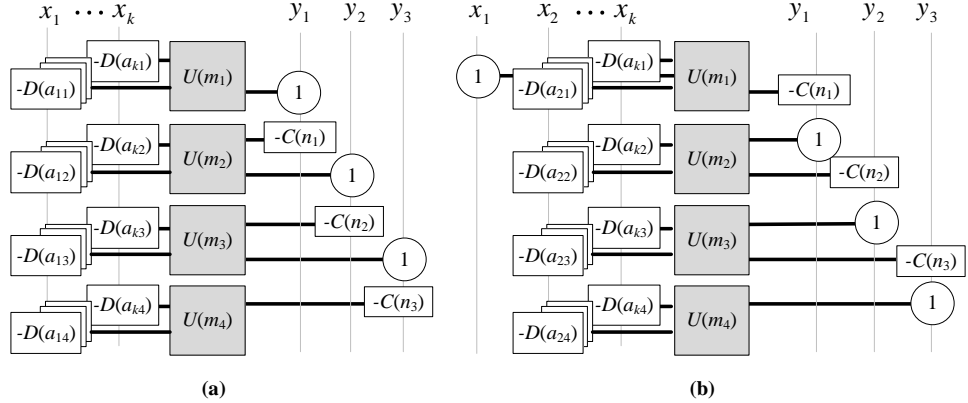

**Fig. 9.** Horn-SAT structure

Figure 9(a) shows the case when the positive variable is used as a chain coupler. In this case, the number system is formed. However, Figure 9(b) shows when the variable is used as a dominant variable candidate. In this case, the positive variable can be only once connected to all decision chains. Therefore, the number system cannot be formed because we cannot make multiple-connected dominant variable candidates of which the feasible value is '1'. Dual-Horn-SAT does not form the number system when the feasible value of a dominant variable is '0' in a tautological sense. ∎

In addition, this paper gives an answer for why random-SAT is so easy.

**Corollary 1. Random-SAT does not form the number system.**

*Proof.* If a variable is used as a chain coupler, random-SAT assigns a similar number of '1' and '-1' to the variable because random-SAT cannot distinguish what is the chain coupler among *n* number of variables. Hence, the exponent of the number system becomes near to 1. ∎

As we know, the polynomial-time algorithm for Horn-SAT is based on the rule of unit propagation. If Horn-SAT has no unit clause, the problem is very easy to solve. If there is no clause including a positive variable, we assign '0' to all variables. If there are one or more positive variables, we first assign '1' to all positive variables and then remove the variables with a negative coefficient in all clauses. If there is no unit clause, we assign '0' to all variables. Then the two instances are satisfiable. However, if the problem changes to determine whether there exists a negative variable that can take '1' as a feasible value, the problem becomes hard as much as the *k*-SAT (*k*>2) because we cannot assign '0' to all negative variables and Horn-SAT forms the number system in the case when the negative variable is a dominant variable by Lemma 3.

As we mentioned, a chain coupler can be connected to two decision chains with a different sign and with a different number of connections. By this property, the chain coupler causes exponentially large coefficients in inequalities of the cutting planes in terms of a linear system, exponentially small distance from the infeasible integral point to the polytope in terms of geometry, and increasing search complexity by multi-branch in terms of a search algorithm such as DPLL (Davis-Putnam-Logemann-Loveland). The result of our study is summarized through the following theorem.

**Theorem 1. We can construct a *k*-SAT (*k*>2) instance where the decision margin decreases to an exponentially small value by the number system that is formed from the calculation of the coefficients in the cutting planes. However, 2-SAT does not form the number system and Horn-SAT forms the number system only when the feasible value of the dominant variable is '0'.**

*Proof.* This theorem is proven by Lemma 1, Lemma2, and Lemma 3. ∎

## 7. Discussion

We showed the natural number system hidden inside the SAT as a clear mathematical expression for the classification of 2-SAT, Horn-SAT, and *k*-SAT (*k*>2) for the first time. We verified that Horn-SAT has an intermediate property between 2-SAT and *k*-SAT (*k*>2) regarding the formation possibility of the natural number system. It is known that NL⊆P and P⊆NP, but unknown if NL=P and P=NP. These questions have been open problems for several decades. As we know, 2SAT is NL-complete, Horn-SAT is P-complete, and *k*-SAT (*k*>2) is NP-complete. Therefore, the formation possibility of the number system becomes a definite supporting evidence for the conjecture that NL⊊P⊊NP. In addition, researchers have treated SAT as a set of polynomial Boolean functions. This is the first study verified that SAT must not be treated as a set of polynomial functions in the process of the linear operations. It was an open problem whether the coefficients of an inequality generated by the cutting plane proof system for a SAT instance are polynomially bounded. Researchers believed that cutting plane proofs with large coefficients are highly non-intuitive[20]. We showed that, against their expectations, the coefficients of an inequality generated by the cutting plane proof system are not polynomially bounded.

**Acknowledgments**

I especially thank **Professor Beom-Hee Lee** from the Department of Electrical and Computer Engineering at Seoul National University. In addition, I thank **Professors Jae-Byung Park** and **Hae-Kyung Cho** for their advice and guidance of this study.